\pdfoutput=1
\documentclass[%
 amsmath,amssymb,
 aps,
]{revtex4}

\usepackage{graphicx}
\usepackage{dcolumn}
\usepackage{bm}

\usepackage{amsmath}
\usepackage[english]{babel}
\usepackage{graphicx,color}
\usepackage{hyperref}

\newcommand{\Z}[2]{\ensuremath{Z^{#1\rightarrow #2}}}
\newcommand{\mess}[3]{\ensuremath{#1^{#2 \rightarrow #3}}}
\newcommand{\neigMeta}[1][e]{\ensuremath{\partial #1^M}}
\newcommand{\neigRea}[1][a]{\ensuremath{\partial #1^R}}

\def\neigGroupI{j \in \partial a^R_{i}}
\def\neigOppositeI{j \in \partial a^R_{\neg i}}
\def\neigIn{j \in \partial a^R_{in}}
\def\neigOut{j \in \partial a^R_{out}}

\def\etheta{e^{\theta}}

\newcommand{\avg}[1]{\left\langle{#1}\right\rangle}

\newcommand{\inn}{{\rm in}}
\newcommand{\outt}{{\rm out}}

\begin{document}


\title{Searching for feasible stationary states in reaction networks\\
by solving a Boolean constraint satisfaction problem}

\author{
A. Seganti$^{1}$, 
F. Ricci-Tersenghi$^{1,2,3}$ and
A. De Martino$^{1,2,4}$
}
\affiliation{%
$^1$ Dipartimento di Fisica, Sapienza Universit\`a di Roma, p.le A.~Moro 2, 00185 Rome (Italy)\\
$^2$ IPCF-CNR, UOS di Roma, Dipartimento di Fisica, Sapienza Universit\`a di Roma (Italy)\\
$^3$ INFN, Sezione di Roma 1, Dipartimento di Fisica, Sapienza Universit\`a di Roma (Italy)\\
$^4$ Center for Life Nano Science@Sapienza, Istituto Italiano di Tecnologia, Viale Regina Elena 291, 00161 Roma (Italy)
}%


\begin{abstract}
We analyze the solutions, on single network instances, of a recently introduced class of constraint-satisfaction problems (CSPs), describing feasible steady states of chemical reaction networks. First, we show that the CSPs generalize the scheme known as Network Expansion, which is recovered in a specific limit. Next, a full statistical mechanics characterization (including the phase diagram and a discussion of physical origin of the phase transitions) for Network Expansion is obtained. Finally, we provide a message-passing algorithm to solve the original CSPs in the most general form.
\end{abstract}

\pacs{Valid PACS appear here}
\maketitle 


\section{Introduction}

The mathematical theory of chemical reaction networks is mainly concerned with issues like the existence, number and stability of fixed points for realistic (e.g. mass-action based) dynamics of continuous state variables representing the concentrations of chemical compounds. In many real-world cases, however, full-fledged dynamical approaches are prevented either by lack of knowledge about kinetic constants or by sheer size considerations. It is so, for instance, for cellular metabolic networks at genome-scale \cite{Beard_2008,Palsson_bookII}. On the other hand, being able to describe viable steady states of the network in terms of coarse-grained, discrete variables might be a useful (albeit less comprehensive) alternative. 

At the simplest level, the operation of networks of chemical reactions can be thought to depend on the availability of reaction substrates and of the enzymes required to process them. In particular, one can think that reactions may occur whenever all of the required substrates are available, which in turn makes the reaction products available, and so on. Likewise, compound availability can be assumed to depend at least on there being an active reaction producing it. Using these ideas, it is possible to associate to a given reaction network (i.e. to a given bipartite graph encoding for reaction/compounds interactions) a set of logical constraints to be satisfied by Boolean state variables indicating whether a reaction is active or inactive and whether a compound is available or not. This type of reasoning has led to the formulation of the Network Expansion (NE) framework \cite{Ebenhoh,Ebenhoh_2004,Ebenhoh_Kruse2008,Handorf_2007}, which aims at quantifying the amount of activity in the network bulk that is generated upon assuming the availability of a seed of compounds. For cellular metabolic networks, the seed usually includes both external species (e.g. nutrients) and internal ones (e.g.  water, currency metabolites like ATP, etc.). In concrete terms, given a seed, one would like to retrieve the pattern(s) of reaction activation/compound availability that are induced via the topology (or, more properly, the stoichiometry) of the reaction network. 

Network Expansion is basically a Boolean Constraint Satisfaction Problem (CSP), possibly one of several types that can be reasonably defined to describe the operation of chemical systems, in analogy with those used in the past to describe other biological mechanisms like transcriptional regulation \cite{Martin,Francois,corre}. Despite their `natural' appeal, however, they have received little attention from the statistical mechanics perspective. Besides the general theoretical interest, extracting biologically significant information from them requires being able to explore (in a controlled way) their very large and rich space of solutions. Devising algorithms that are able to carry out this task, even for the basic NE scheme, is however far from trivial.

Recently \cite{Article_popDyn}, we have proposed a class of CSPs inspired by the so-called constraint-based models for flux analysis \cite{Palsson_book} which aims at describing the space of configurations of a chemical reaction network through minimal Boolean feasibility constraints. These CSPs have been defined on random reaction networks (RRN), i.e. bipartite graphs (the two classes of nodes corresponding to `reactions' and `reagents', respectively) characterized by the parameters $\lambda$, representing the mean of the Poisson distributed degrees of metabolites,
and $q$ (resp. $1-q$), giving the probability that a reaction has two (resp. one) input or output compounds. 
The structure of a RRN is described by a $M\times N$ connectivity matrix $\widehat{\xi}$, with $\xi_i^m\in\{1,0,-1\}$ depending on whether compound $m\in\{1,\ldots,M\}$ is a substrate ($\xi_i^m=-1$), a product ($\xi_i^m=1$) or is not involved ($\xi_i^m=0$) in reaction $i\in\{1,\ldots,N\}$. In this kind of networks, a {\it nutrient} is a compound with in-degree $0$, while a {\it sink} has out-degree $0$. Denoting by $\nu_i\in\{0,1\}$ (inactive/active) the state associated to reaction $i$ and by $\mu_m\in\{0,1\}$ (unavailable/available) that associated to compound $m$, for a given RRN feasible assignments $(\boldsymbol{\mu}=\{\mu_m\},\boldsymbol{\nu}=\{\nu_i\})$ are defined to be such that $\Gamma_m=1~\forall m$ and $\Delta_i=1~\forall i$, where
\begin{gather}
\Gamma_m = \delta_{\mu_m,0}\delta_{x_m,0} (\delta_{y_m,0})^\alpha + \delta_{\mu_m,1}(1-\delta_{x_m,0}) (1-\delta_{y_m,0})^{\alpha} \label{constraint_meta}\\
\Delta_i =\delta_{\nu_i,0}+\delta_{\nu_i,1}\prod_{m \in \partial i_{\inn}}\mu_m~~, \label{constraint_rea}
\end{gather}
$\partial i_{\inn}$ is the set of substrates of reaction $i$, $\alpha\in\{0,1\}$ is a fixed parameter, and
\begin{equation}
x_m\equiv \sum_{i\in\partial m_{\inn}} \nu_i\;\quad ~~~~~\text{and}~~~~~
y_m\equiv \sum_{i\in\partial m_{\outt}} \nu_i~~,
\end{equation}
with $\partial m_{\inn}$ (resp. $\partial m_{\outt}$) the set of reactions producing (resp. consuming) chemical species $m$. Condition (\ref{constraint_rea}) says that reactions can always be inactive, while they can activate only if all input compounds are available; likewise, condition (\ref{constraint_meta}) allows for a compound $m$ to be available if at least one reaction produces it (for $\alpha=0$) or if  at least one reaction produces it and one consumes it (for $\alpha=1$). As explained in detail in \cite{Article_popDyn}, the case $\alpha=0$ (`Soft Mass Balance' or Soft-MB) describes steady states that allow for a net production of compounds, while the case $\alpha=1$ (`Hard Mass Balance' or Hard-MB) corresponds, in this coarse-grained view, to a fully mass-balanced scenario. 

Once the topology of the reaction network, encoded in an adjacency matrix $\widehat{\xi}$, is given, the setup presented in \cite{Article_popDyn} aims at retrieving Boolean patterns of activity of reactions (or of metabolite availabilities) induced by the fact that a certain seed of metabolites (in our case, formed by nutrients only) is available from the outset. The cavity-based population dynamics technique developed in \cite{Article_popDyn} allows in particular to sample configurations $(\boldsymbol{\mu},\boldsymbol{\nu})$ with a probability given by
\begin{equation}
P(\boldsymbol{\mu},\boldsymbol{\nu}) \propto \prod_{m=1}^M \Gamma_m \prod_{i=1}^N \Delta_i e^{\theta \nu_i}\;,
\label{meas}
\end{equation}
with a `chemical potential' $\theta$ that allows to select states according to the overall number of active processes and, in turn, compute network ensemble-averaged quantities. This study has revealed a rich phase structure characterized by hysteresis, which might potentially hinder the retrieval of individual solutions.

Here we extend the previous analysis in a direction hopefully more useful for applications to quantitative biology (which will be our next step), by searching for solutions to the above CSP on a {\it given} reaction network. In this context we are going to discuss the statistical properties of the solutions found by several search methods (including NE) and to introduce an improved decimation-based technique. Furthermore, we will compare the statistical properties of the solutions found on the single instance with the ensemble-averaged results obtained in \cite{Article_popDyn}. The details of the cavity theory on which our algorithms are based, as well as of the algorithms themselves (Belief Propagation complemented by decimation), are reported in the Appendices.

\section{Network Expansion revisited}

\subsection{The problem}
\label{sec:mean_field}
\begin{figure}
\includegraphics[height=5cm]{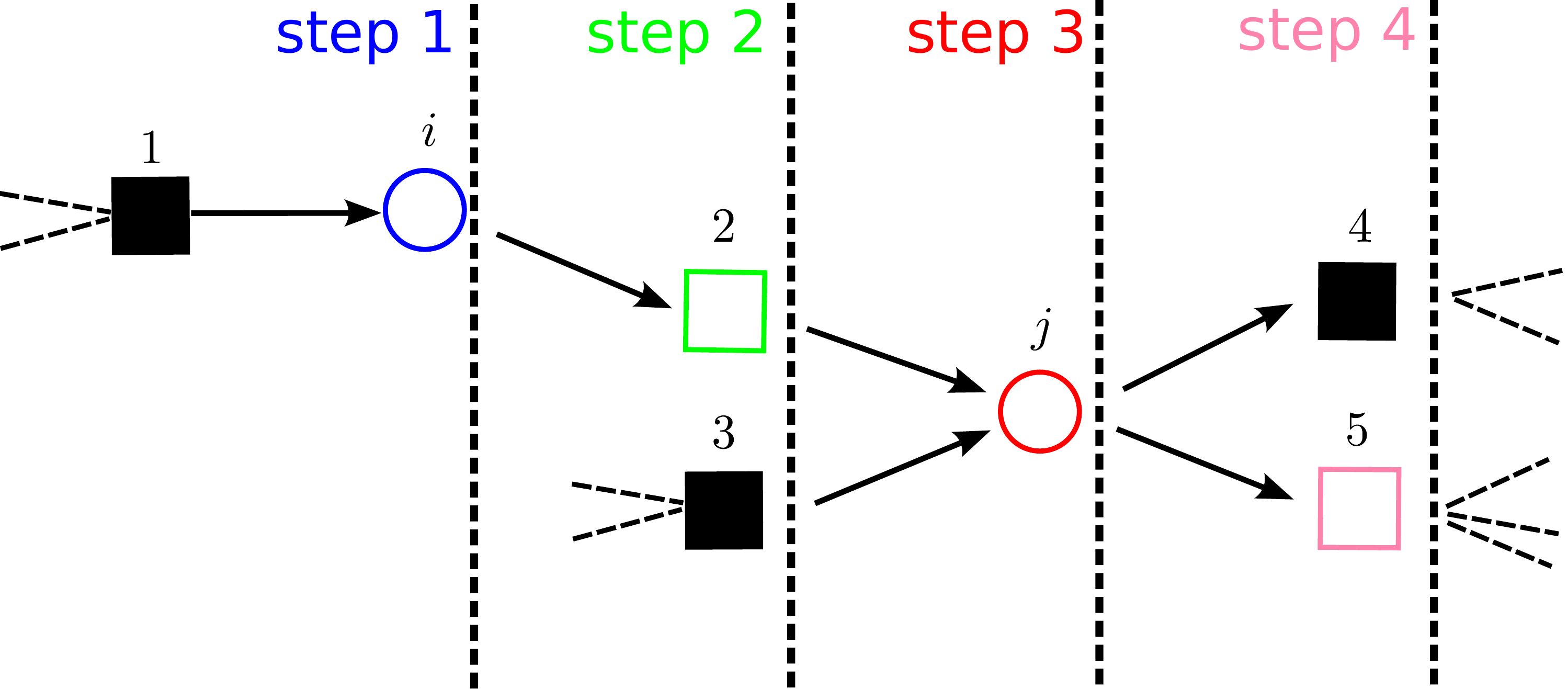}
\caption{\label{steps_NE}Sketch of four steps of the Propagation of External Inputs (PEI) algorithm (serving as the basis of the Network Expansion method \cite{Ebenhoh}). Black squares represent compounds initially available. In step 1, reaction $i$ is activated by virtue of the availability of compound $1$; in step 2, metabolite $2$ becomes available by virtue of the activation of reaction $i$; in step 3, reaction $j$ activates as both $2$ and $3$ are available; and so on.}
\end{figure}

The basic idea behind NE is that, given a seed compound (e.g. a nutrient), a reaction can (and will) activate when all its substrates are available (AND-like constraint), whereas a compound will be available if at least one of the reactions that produce it is active (OR-like constraint). The numerical procedure of NE transfers the information about the availability of certain metabolites across the network links, as explained pictorially in Fig. \ref{steps_NE}. We shall term this type of process a Propagation of External Inputs (PEI). 

It is simple to understand that, as soon as the reaction network departs from a linear topological structure, the propagation will likely stop after a small number of steps unless the availability of additional compounds is invoked. Indeed, in Network Expansion PEI is aided by the assumption that highly connected metabolites like water are abundant. Because  of its intuitive appeal, it is useful to analyze briefly the properties of PEI in somewhat more detail.

One can write down equations for the probability $\avg{\nu_i}$ that reaction $i$ will be active and for the probability $\avg{\mu_m}$ that metabolite $m$ will be available by simply considering that, under PEI in a given network, a reaction can activate when all of its inputs are available and a metabolite becomes available when at least one reaction is producing it. This implies that
\begin{gather}
\avg{\nu_i}=\prod_{n \in \partial i_{\inn}} \avg{\mu_n}~~, \label{MF_thetaINF_rea} \\
1 - \avg{\mu_m} = \prod_{k \in \partial m_{\inn}}(1-\avg{\nu_k}) \label{MF_thetaINF}~~.
\end{gather}
To prove the link between PEI and the CSPs defined above, note that, using the definition (\ref{meas}) one can easily compute the mean values
\begin{gather}
\avg{\nu_j}=\frac{\sum\limits_{\boldsymbol{\mu},\boldsymbol{\nu}} \nu_j \prod\limits_{m=1}^M \Gamma_m \prod\limits_{i=1}^N \Delta_i e^{\theta \nu_i}}{\sum\limits_{\boldsymbol{\mu},\boldsymbol{\nu}}  \prod\limits_{m=1}^M \Gamma_m \prod\limits_{i=1}^N \Delta_i e^{\theta \nu_i}}~~,\\
\avg{\mu_n}=\frac{\sum\limits_{\boldsymbol{\mu},\boldsymbol{\nu}} \mu_n \prod\limits_{m=1}^M \Gamma_m \prod\limits_{i=1}^N \Delta_i e^{\theta \nu_i}}{\sum\limits_{\boldsymbol{\mu},\boldsymbol{\nu}}  \prod\limits_{m=1}^M \Gamma_m \prod\limits_{i=1}^N \Delta_i e^{\theta \nu_i}}~~
\end{gather}
Under the Mean Field Approximation, we can set
\begin{gather}
 \Gamma_m(\mu_m,\{\nu_i\})=\Gamma_m(\mu_m,\{\langle \nu_i \rangle \})~~,\\
\Delta_i(\nu_i,\{\mu_m\})=\Delta_i(\nu_i,\{\avg{\mu_m}\})~~,
\end{gather}
which in turn implies
\begin{gather}
\label{punto_fisso_theta_fin}
\langle \nu_i \rangle=\frac{\etheta \prod\limits_{n \in \partial i_{\inn}}\langle\mu_n\rangle}{1+\etheta \prod\limits_{n \in \partial i_{\inn}}\langle\mu_n\rangle} \\
\langle \mu_m \rangle=1-\prod_{k \in \partial m_{\inn}}(1-\langle\nu_j\rangle)~~.
\end{gather}
In the limit $\theta\rightarrow \infty$ we have
\begin{gather}
\avg{\nu_i}= 
\begin{cases}
1 &\text{if }\prod\limits_{n \in \partial i_{\inn}}\langle\mu_n\rangle = 1~~, \\ 
0 &\text{if } \prod\limits_{n \in \partial i_{\inn}}\langle\mu_n\rangle = 0~~.
\end{cases}
\end{gather}
So that equations (\ref{MF_thetaINF_rea}) and (\ref{MF_thetaINF}) are recovered. In other terms, PEI is the Mean Field Approximation at $\theta\to\infty$ of the CSPs considered in \cite{Article_popDyn}.

\begin{figure}
\includegraphics[width=10cm]{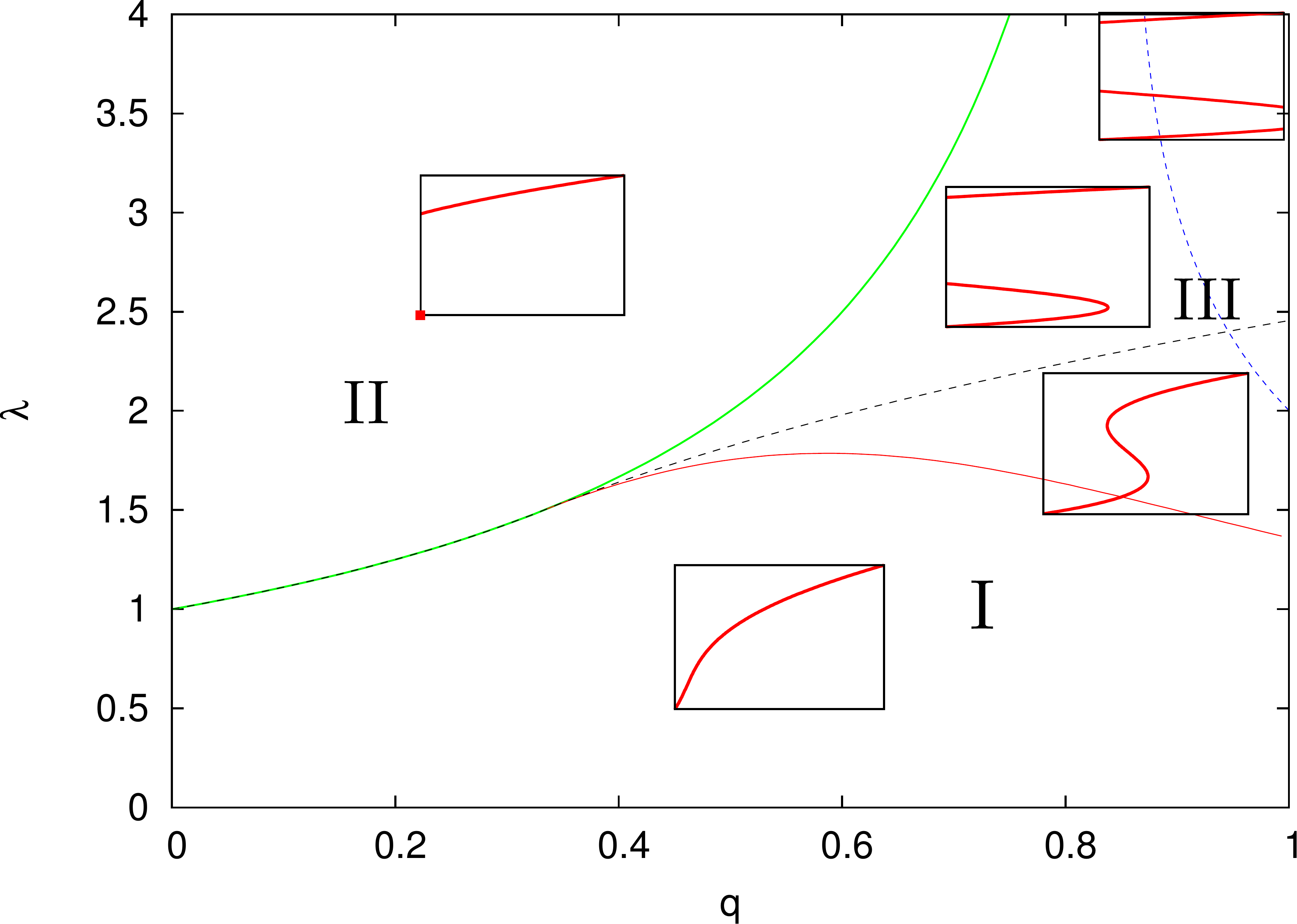}
\caption{(color online) Phase diagram obtained by propagation of external inputs (PEI) on RRN in the $(q,\lambda)$ plane. The insets display the curves $\gamma^\star$ versus $\rho_{\inn}$ obtained in the different sectors; all lines are analytical. See text for details.}
\label{diagramma_fase_theta_infty}
\end{figure}

It is simple to derive analytically the phase diagram of PEI in the ensemble of RRN defined in \cite{Article_popDyn}. The probability that a metabolite is available is
\begin{equation}
\gamma=\overline{\avg{\mu_m}}~~,
\end{equation}
where the over-bar denotes an average over the network realizations. Using (\ref{MF_thetaINF_rea}) and (\ref{MF_thetaINF}) one sees that
\begin{gather*}
\gamma =e^{-\lambda}\rho_{\inn}+\sum_{k_m \geq 1}\mathit{D}_{M}(k_m)\left(1-\prod_{j=1}^{k_m}(1-\overline{\avg{\nu_j}})\right)~~,
\end{gather*}
where we have assumed that nutrients (fractionally given by roughly $e^{-\lambda}$ nodes) have a fixed probability $\rho_{\inn}$ of being available and where $D_M(k)=e^{-\lambda}\lambda^k/k!$ is the distribution of metabolite in- (and out-)degrees. In turn, this gives
\begin{gather}
\gamma=e^{-\lambda}\rho_{\inn}+1-e^{-\lambda\tau}~~,
\label{eq_punto_fisso_M}
\end{gather}
where $\tau=\overline{\avg{\nu_i}}$ is the probability that a reaction is active, which, recalling that the in- and out-degrees of reactions are distributed according to $D_R(d)=q\delta_{d,2}+(1-q)\delta_{d,1}$, satisfies (within a Mean-Field Approximation)
\begin{gather}
\tau=\overline{\prod_{b\in\partial i_{\inn}} \avg{\mu_n}}=(1-q)\gamma+q\gamma^2~~.
\label{eq_punto_fisso_R}
\end{gather}
Putting things together, $\gamma$ is seen to satisfy the condition
\begin{equation}
\gamma=e^{-\lambda}\rho_{\inn}+1-\exp[-\lambda((1-q)\gamma+q\gamma^2)]~~,
\label{eq_punto_fisso_theta_infty}
\end{equation}
which can be solved for $\gamma$ upon changing the values of $\rho_{\inn}$, $q$ and $\lambda$. The resulting phase diagram in the $(q,\lambda)$ plane, based on the behaviour of the solution $\gamma^\star(\rho_\inn)$, is displayed in Figure \ref{diagramma_fase_theta_infty}.

Three regions can be distinguished. In region I,  Equation (\ref{eq_punto_fisso_theta_infty}) has a unique solution and $\gamma^\star$ is a monotonously increasing function of $\rho_\inn$ (note that $\gamma^\star=0$ is always a solution when $\rho_\inn=0$). Outside region I, the curve $\gamma^\star$ vs $\rho_\inn$ displays an inflection point. If the point lies outside the interval $[0,1]$ (for both $\gamma$ and $\rho_\inn$) then (\ref{eq_punto_fisso_theta_infty}) has a unique non-zero solution for $\rho_\inn>0$ and two different solutions at $\rho_\inn=0$ (region II). In region III, instead, a range of values of $\rho_\inn$ exists where three distinct solutions (with different values of $\gamma$) of (\ref{eq_punto_fisso_theta_infty}) occur. This sector can be further divided according to the number of solutions found for $\rho_\inn=0$ and $\rho_\inn=1$. The black dashed line marks the boundary between phases with, respectively, one and three solutions for $\rho_\inn=0$ while the dashed blue line separates the region with one and three solutions for $\rho_\inn=1$.

\begin{figure}
\includegraphics[width=12cm]{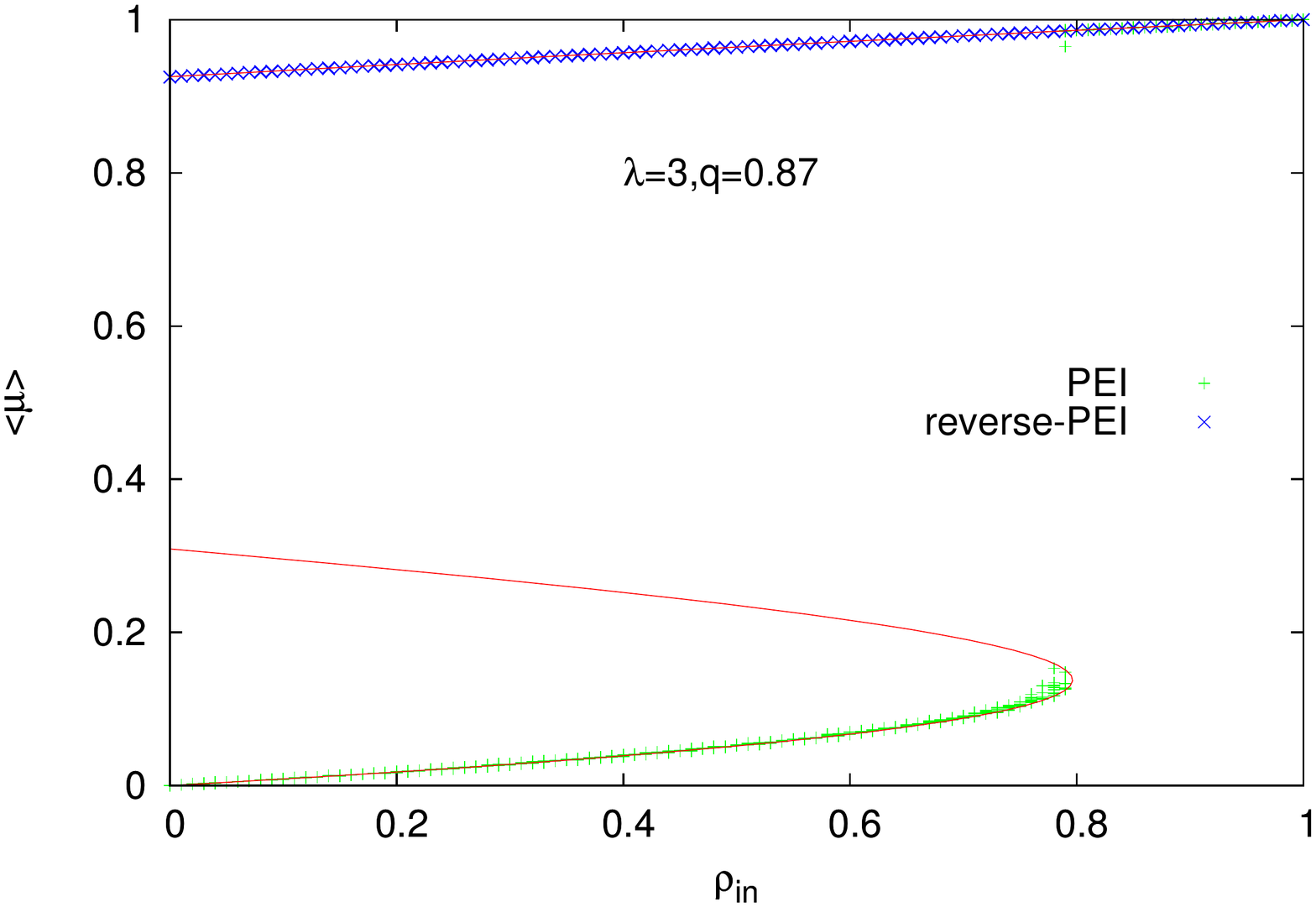}
\caption{Theoretical solution of Equation (\ref{eq_punto_fisso_theta_infty}) (solid line) versus $\rho_{\inn}$, together with the results obatined by PEI and reverse-PEI.}
\label{PEI_sampling}
\end{figure}

For any fixed $\rho_\inn$, whenever solutions with different values of $\avg{\mu}$ coexist, those with the smallest $\avg{\mu}$ can be retrieved by straightforward PEI starting from a configuration where no metabolite is available except for nutrients. Solutions with larger $\avg{\mu}$, on the other hand, can be found by `reverse-PEI'. In this procedure a configuration where internal metabolites are all available and nutrients are fixed with probability $\rho_{\inn}$ is initially selected, and then a solution is found by enforcing the constraints in an iterative way. The results for both procedures are presented in Figure \ref{PEI_sampling} for $\lambda=3$ and $q=0.87$ (deep into region III in Figure \ref{diagramma_fase_theta_infty}).

\subsection{Origin of the phase transition within the Mean-Field Approximation in PEI}
\label{percolation_transition}

We show here that, as might have been intuitive, the phase transitions occurring in PEI (see Figure \ref{diagramma_fase_theta_infty}) are, from a physical viewpoint, of a percolation type.

\begin{figure}
\includegraphics[width=12cm]{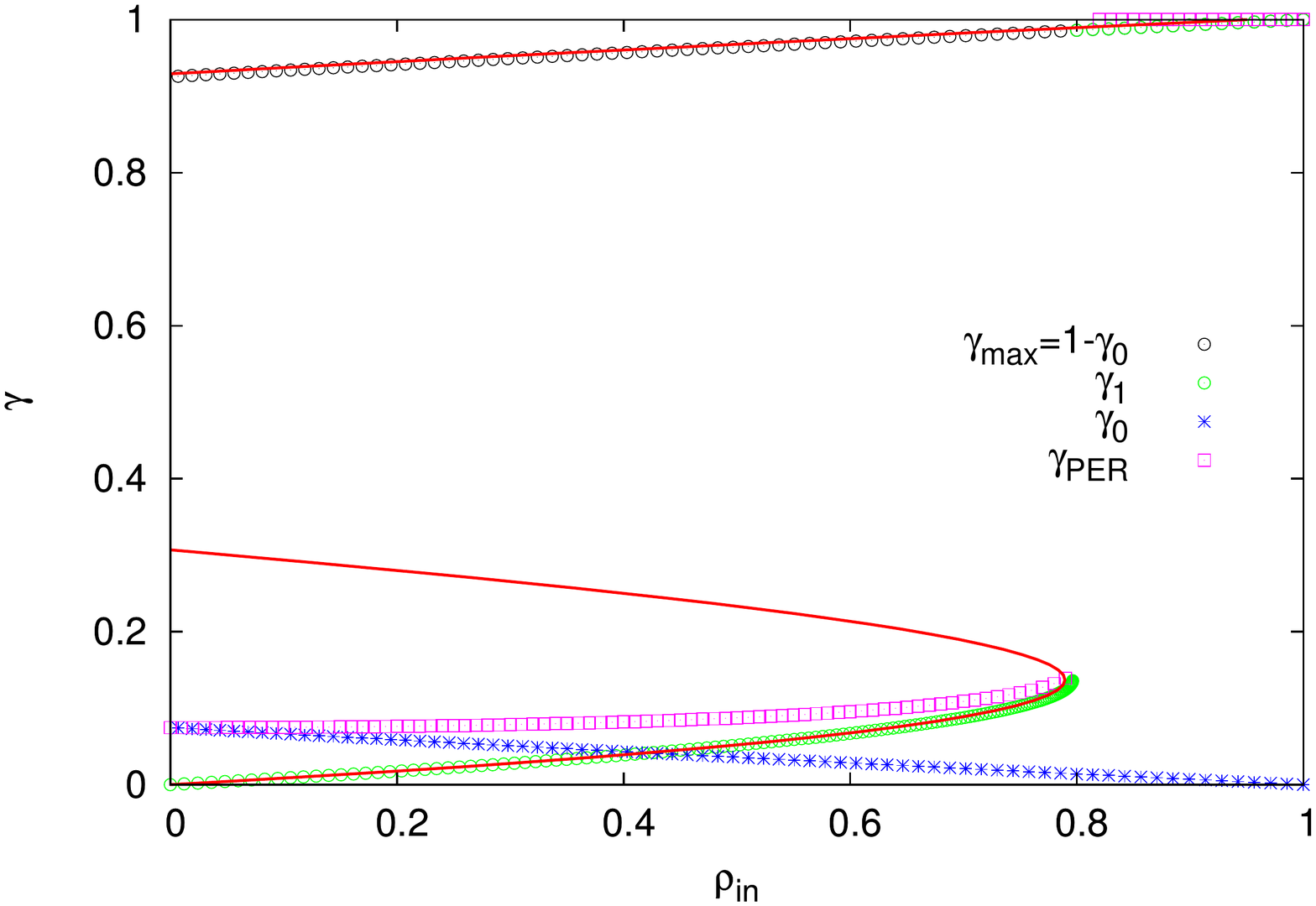}
\caption{Weights of the different components of the PER core for a graph with $\lambda=3$ and $q=0.87$. The red line corresponds to the solution of equation (\ref{eq_punto_fisso_theta_infty}).\label{PERcore_fig}}
\end{figure}

To analyze the effectiveness of PEI, we start by identifying the so-called Propagation of External Regulation (PER) Core of the system \cite{Leone_Zecchina}, that is the sub-network obtained by fixing the nutrient availability (with probability $\rho_\inn$) and then propagating this information inside the network. In this way, some variables will be assigned a definite value (either $1$ or $0$). At convergence, a fraction $\gamma_1$ (resp. $\gamma_0$) of metabolites will be fixed to $1$ (resp. $0$), while a fraction $\tau_1$ (resp. $\tau_0$) of reactions will be fixed to $1$ (resp. $0$). One easily sees that, at the fixed point, the following equations hold:
\begin{gather}
1-\tau_0=q(1-\gamma_0)^2+(1-q)(1-\gamma_0)~~,\\
\gamma_0=\sum\limits_{k\neq 0}\mathit{D}_M(k)\tau_0^k+(1-\rho_{\inn})D_M(0)~~,\\
    \tau_1=q\gamma_1^2+(1-q)\gamma_1~~, \\
    1-\gamma_1=\sum\limits_{k\neq 0}\mathit{D}_M(k)(1-\tau_1)^k+(1-\rho_{\inn})D_M(0)~~.
\end{gather}
In turn, one obtains
\begin{gather}
\tau_0=1-q(1-\gamma_0)^2-(1-q)(1-\gamma_0)~~,\\ 
\gamma_0=e^{-\lambda}e^{\lambda\tau_0}-\rho_{in}e^{-\lambda}~~,\\
    \tau_1=q\gamma_1^2+(1-q)\gamma_1~~,\\ 
    \gamma_1=1-e^{-\lambda\tau_1}+\rho_{in}e^{-\lambda}~~.   
\end{gather}
Unsurprisingly, the equations for $\gamma_1$ and $\tau_1$ take us back to (\ref{eq_punto_fisso_theta_infty}). On the other hand, the fraction of metabolites in the PER core is given by
\begin{equation}
\gamma_{PER}=\gamma_1+\gamma_0~~. 
\end{equation}
Hence the fraction of metabolites that are not fixed by propagating nutrient availability is given by $1-\gamma_{PER}$, and the maximum achievable availability for metabolites (that we will often call `magnetization' in the following using a statistical physics jargon) is given by $\gamma_{max}=1-\gamma_0$. 
Figure \ref{PERcore_fig} displays the different contributions for a specific choice of the parameters, together with the corresponding solution of Eq. (\ref{eq_punto_fisso_theta_infty}).

The excellent agreement of $\gamma_1$ with the analytical line for the feasible values of the magnetization suggests that straightforward PEI will be able to recover solutions with lower magnetization when the latter coexist with high-magnetization solutions. On the other hand, the highest magnetizations coincide, expectedly, with the largest achievable average metabolite availability. Finally, depending on the value of $\lambda$ and $q$, one obtains a single solution when no PER core exists, and two solutions (with magnetizations $\gamma_{max}$ and $\gamma_1$) in presence of a PER core. Hence the transition is a typical percolation transition between a phase in which the internal variables are trivially determined by the nutrients (in absence of a PER core) to one in which the internal variables are not univocally determined (in presence of a PER core).

\section{Solutions on individual networks by Belief Propagation and decimation}

\begin{figure}
\includegraphics[scale=0.3]{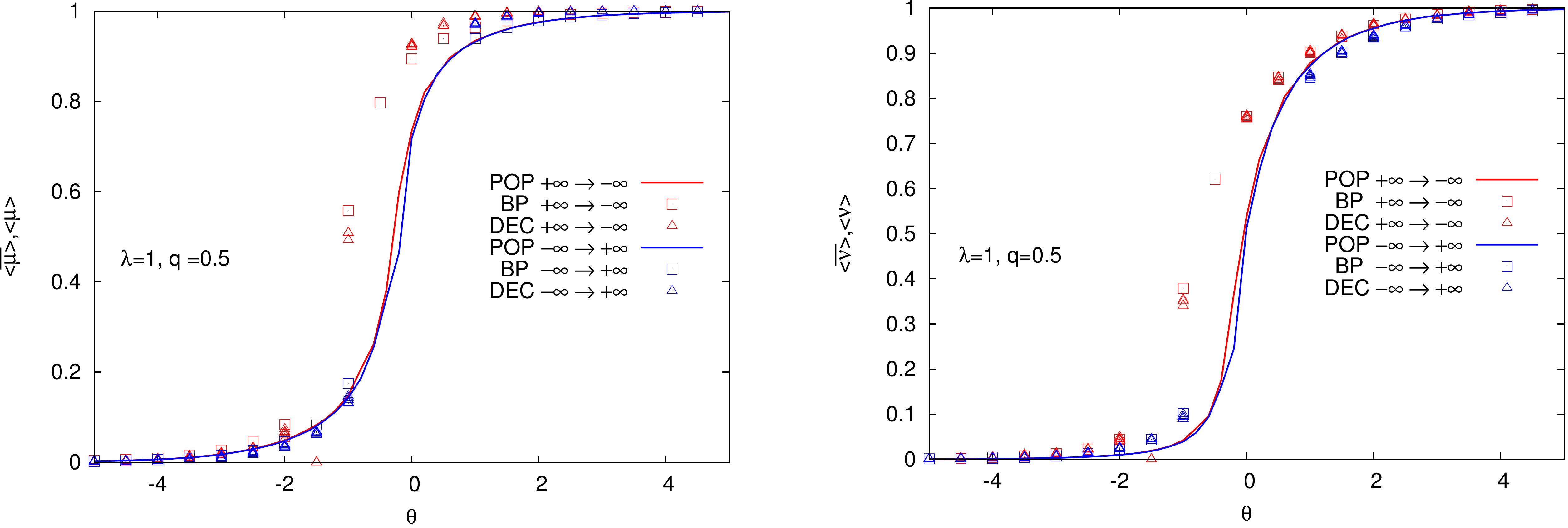}
\caption{{\bf Soft-MB} for $\lambda=1$, $q=0.5$ and $\rho_{\inn}=1$. {\bf Left}: average fraction of available metabolites, $\avg{\mu}$ ($\overline{\avg{\mu}}$ for population dynamics) versus $\theta$. {\bf Right}: average fraction of active reactions, $\avg{\nu}$ ($\overline{\avg{\nu}}$ for population dynamics) versus $\theta$.\label{VN_l1_q05}}
\end{figure}

\begin{figure}
\includegraphics[scale=0.3]{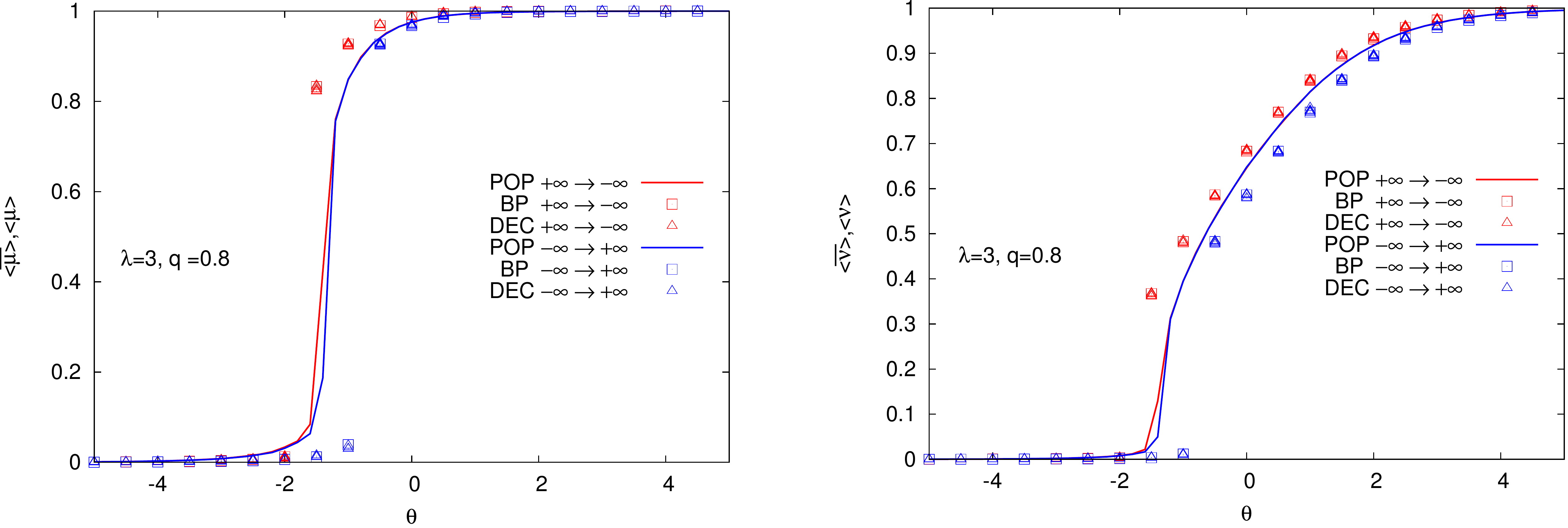}
\caption{{\bf Soft-MB} for $\lambda=3$, $q=0.8$ and $\rho_{\inn}=1$. {\bf Left}: average fraction of available metabolites, $\avg{\mu}$ ($\overline{\avg{\mu}}$ for population dynamics) versus $\theta$. {\bf Right}: average fraction of active reactions, $\avg{\nu}$ ($\overline{\avg{\nu}}$ for population dynamics) versus $\theta$.\label{VN_l3_q08}}
\end{figure}

\begin{figure}
\center
\includegraphics[scale=0.3]{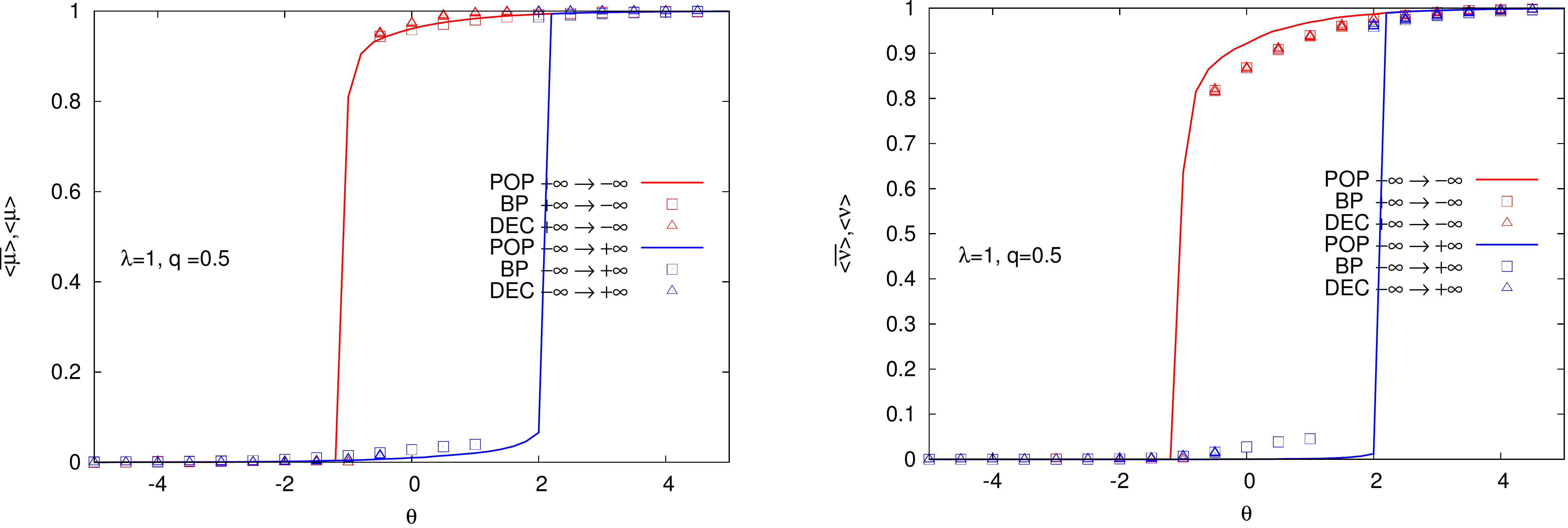}
\caption{{\bf Hard-MB} for $\lambda=1$, $q=0.5$ and $\rho_{\inn}=1$. {\bf Left}: average fraction of available metabolites, $\avg{\mu}$ ($\overline{\avg{\mu}}$ for population dynamics) versus $\theta$. {\bf Right}: average fraction of active reactions, $\avg{\nu}$ ($\overline{\avg{\nu}}$ for population dynamics) versus $\theta$.\label{FBA_l1_q05}}
\end{figure}

\begin{figure}
\center
\includegraphics[scale=0.3]{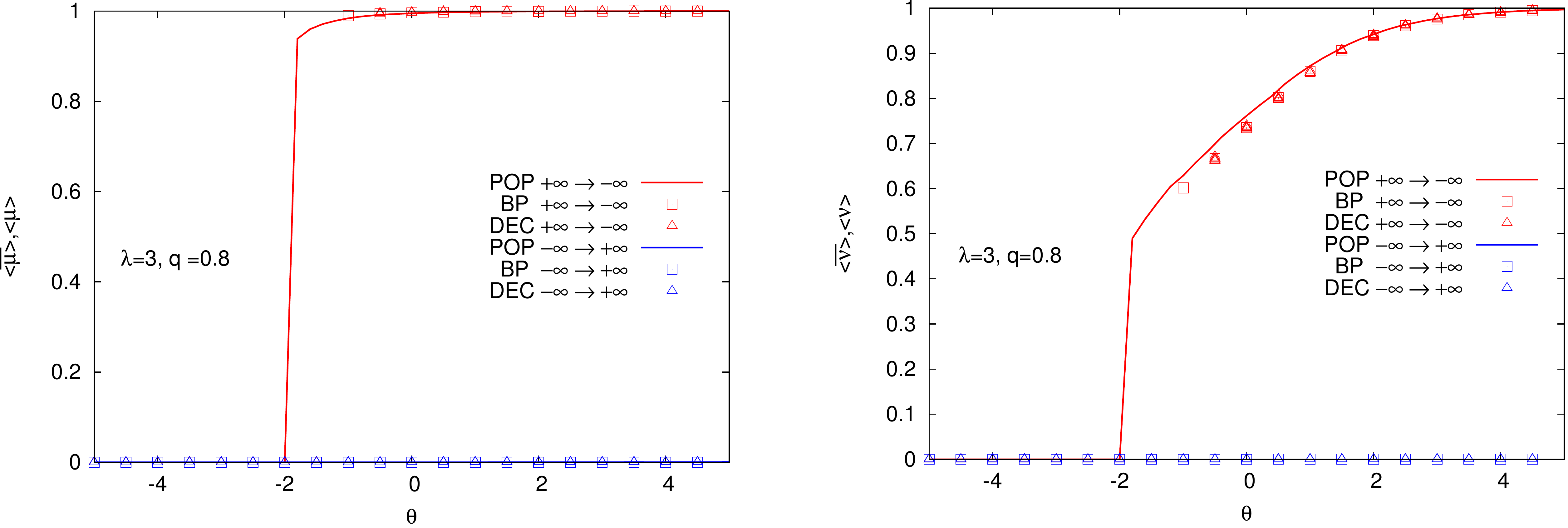}
\caption{{\bf Hard-MB} for $\lambda=3$, $q=0.8$ and $\rho_{\inn}=1$. {\bf Left}: average fraction of available metabolites, $\avg{\mu}$ ($\overline{\avg{\mu}}$ for population dynamics) versus $\theta$. {\bf Right}: average fraction of active reactions, $\avg{\nu}$ ($\overline{\avg{\nu}}$ for population dynamics) versus $\theta$.\label{FBA_l3_q08}}
\end{figure}

We turn now to the analysis of the Soft-MB and Hard-MB CSPs (\ref{constraint_meta}) and (\ref{constraint_rea}) for general $\theta$. In essence, we have derived the cavity equations for the CSPs, presented in Appendix \ref{cavity_equation}, and used the Belief Propagation (BP) algorithm discussed in Appendix \ref{BP_explanation} to compute the statistics of solutions on single instances of RRNs. Next, in order to obtain  \textit{individual configurations} of variables that satisfy our CSPs, we resorted to the decimation scheme presented in Appendix \ref{decimation_explanation}. Results are presented in Figures \ref{VN_l1_q05} and \ref{VN_l3_q08} for Soft-MB ($\alpha=0$) and in Figures \ref{FBA_l1_q05} and \ref{FBA_l3_q08} for Hard-MB ($\alpha=1$). Results from Belief propagation, labeled as `BP', are compared with results retrieved by the population dynamics algorithm developed in \cite{Article_popDyn} (labeled `POP' and corresponding to the ensemble average) and with the decimation results (labeled `DEC'). In \cite{Article_popDyn}, the solution space was explored by two different protocols, which we also use here: by reducing $\theta$ starting from a large positive value ($+\infty \rightarrow -\infty$ in the Figure legends) and by doing the reverse ($-\infty \rightarrow +\infty$ in the Figure legends). If the decimation scheme does not converge, the corresponding point is absent.

It is  clear that decimation generically fails to converge close to the transitions both in the Soft-MB and, more severely, in the Hard-MB case. Apart from this, the three methods give results that are in remarkable qualitative agreement, including  the ability to describe discontinuities in $\langle \mu \rangle$ and $\langle \nu \rangle$ upon varying $\theta$. It is noteworthy that many different configurations appear to be feasible. These configurations are spread over a broad range of densities, especially in the Soft-MB case. So our method based on BP and decimation is able to sample the solution space by just varying a single parameter (the chemical potential $\theta$ in the present case), even in cases when only ``extremal'' solutions seem to satisfy the CSP for metabolite nodes (as e.g. in the left panel in Fig.~\ref{FBA_l3_q08}) while the density of active reaction is varying in a more continuous manner (see the right panel in the same figure).

\begin{figure}
\center
\includegraphics[scale=0.3]{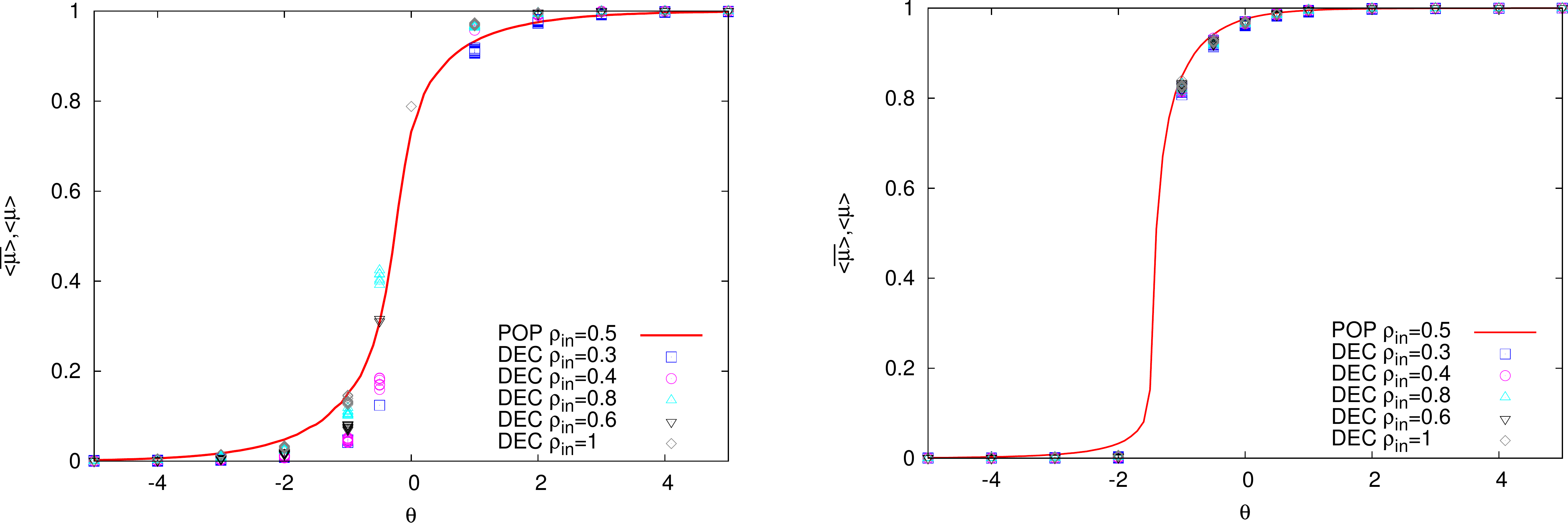}
\caption{{\bf Soft-MB}:  behaviour of the average fraction of available metabolites, $\avg{\mu}$ ($\overline{\avg{\mu}}$ for population dynamics) for $\lambda=1$ and $q=0.5$ (Left) and $\lambda=3$ and $q=0.8$ (Right) at various $\rho_{\inn}$.\label{VN_AllRhoIn}}
\end{figure}

\begin{figure}
\center
\includegraphics[scale=0.3]{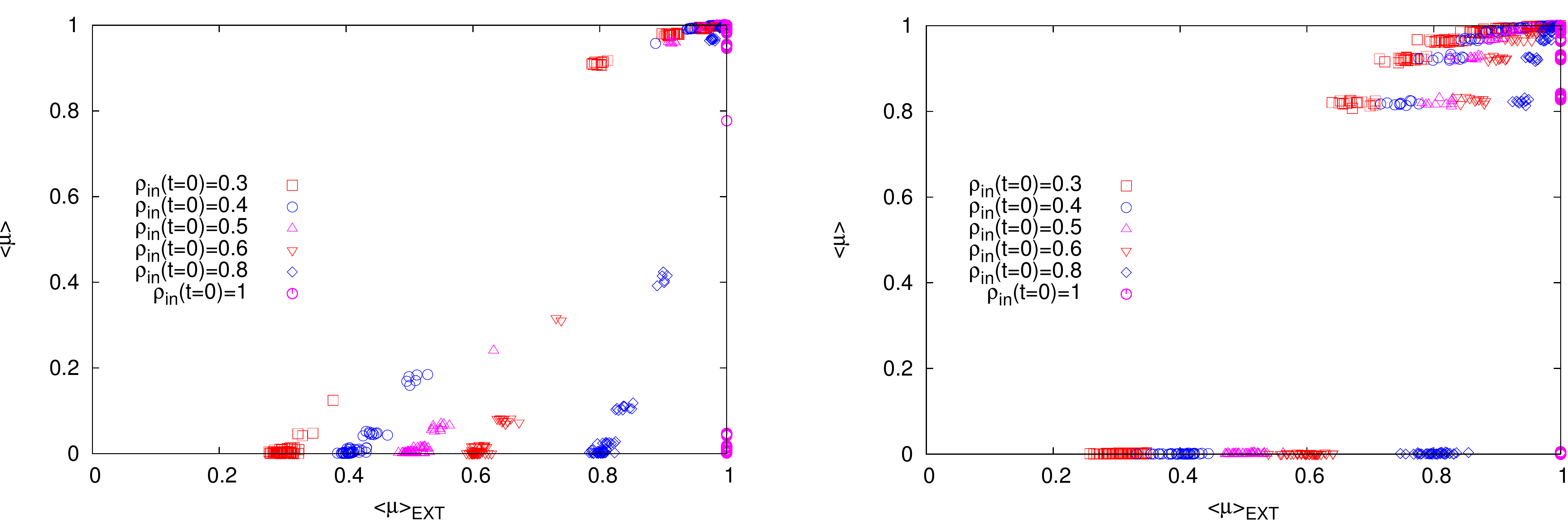}
\caption{{\bf Soft-MB}: Behaviour of $\avg{\mu}$ vs $\avg{\mu}_{EXT}$ for various $\rho_{\inn}$, for $\theta=(-5,-4.5,..,4.5,5)$ and for $\lambda=1$ and $q=0.5$ (Left) and $\lambda=3$ and $q=0.8$ (Right). \label{VN_rhoInVSMagnIn}}
\end{figure}

As detailed in Appendix \ref{cavity_equation}, during decimation nutrients must be treated with special care. This is because the prior assignment of availability for each nutrient (which, as said above, follows a probabilistic rule with parameter $\rho_{\inn}$) does not always coincide, after decimation, with the frequency with which the nutrient is available in the final assignments (i.e. the actual solutions retrieved), which we denote as $\avg{\mu}_{EXT}$. We analyze the relation between the average magnetization of reactions and metabolites and both $\rho_{\inn}$ and $\avg{\mu}_{EXT}$ in Figures \ref{VN_AllRhoIn} and \ref{VN_rhoInVSMagnIn}. We first note that in this way we are able to obtain solutions at various $\avg{\mu}_{EXT}$ clearly different from the corresponding values of $\rho_{\inn}$. Moreover, solutions are rather stable against changes in $\rho_{\inn}$, as is to be expected expected in random networks, at least for the Soft-MB problem. Hard-MB presents however more difficulties (not shown): because it typically admits solutions with either very high or very low magnetization, it turns out to be hard to obtain solutions with $\avg{\mu}_{EXT}\neq 1$, apart from the trivial case when the whole network is inactive.

\begin{figure}
\center
\includegraphics[scale=0.3]{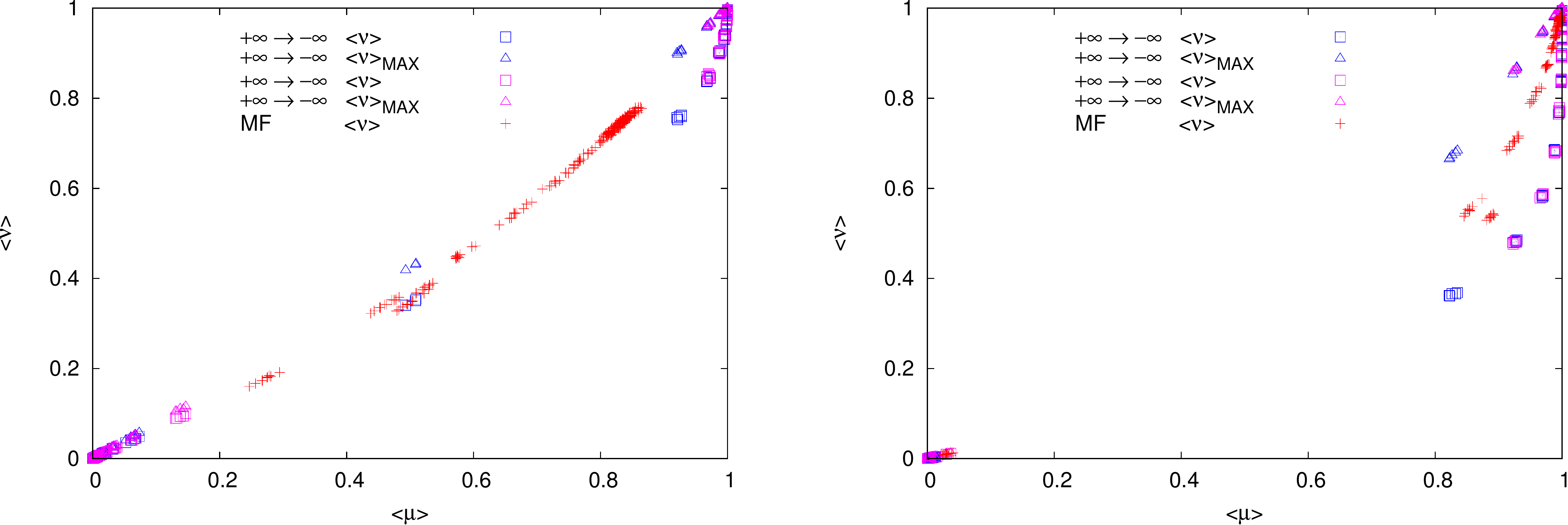}
\caption{Plot of $\avg{\mu}$ versus $\avg{\nu}$ for $\lambda=1$ and $q=0.5$ (Left) and $\lambda=3$ and $q=0.8$ (Right). \label{MF_complete_cmp}}
\end{figure}

Finally we would like to compare the solutions of the complete problem to the solutions obtained using the Mean Field Approximation (MF) presented in the previous Section. Indeed in Section \ref{sec:mean_field} we showed how to obtain solutions for the MF problem at $\theta\rightarrow\infty$ using the PEI or reverse-PEI procedure. However in order to compare the two approaches, MF solutions at all $\theta$ have to be studied. This can be done by searching for solutions to the MF equations (\ref{punto_fisso_theta_fin}) at finite $\theta$,  and then using the same decimation procedure presented in Appendix \ref{decimation_explanation}. It is  important to notice, however, that in the MF case both BP and decimation algorithm can be written in a simpler form as only one message per variable is needed; furthermore, for $\theta\rightarrow\infty$, BP and decimation together behave exactly as a Warning Propagation Algorithm \cite{Braunstein_2005_survey}.

Thus in Figure \ref{MF_complete_cmp} we present the magnetizations $\langle \mu \rangle$ and $\langle \nu \rangle$ for solutions obtained using both MF and the complete problem for various $\theta\in (-\infty,\infty)$. Here MF solutions are represented by crosses, while squares represent the densities of the solutions obtained by the complete decimation algorithm. The $\avg{\nu}$ density in the latter solutions seems to be always smaller than in MF. However we have to remind that our CSP allows for configurations where a reaction is inactive even if all its neighbouring metabolites are present. In these cases, the reaction can be switched on without violating any constraint. The data marked $\avg{\nu}_{MAX}$ in Fig.~\ref{MF_complete_cmp} have been obtained by switching on all possible reaction without changing the configuration of metabolites. This is the upper bound for the reaction activity in the complete problem. 

From data shown in Figure \ref{MF_complete_cmp} it is clear that the complete problem allows for a wider variability in the values of $\avg{\mu}$ and $\avg{\nu}$. Moreover the solutions sampled at MF level are a subpart of the solutions found in the complete problem. Nevertheless, the MF approach may be useful for real networks, being simpler to solve and hence much faster to sample.

\section{Conclusions}

Stationary states of chemical reaction networks can be often described in a compact way through the information regarding reaction activity/inactivity and reagent availability/inavailability. In these conditions, Boolean CSPs provide a framework to describe feasible operation states of chemical reaction networks. The problem posed by sampling their solution space (even for an individual network, as discussed here) is however substantial. We have presented an efficient computational method to generate solutions for a  class of CSPs inspired by constraint-based models of cell metabolism. Extending previous work concerned with ensemble properties, we have focused here on characterizing the solution space for single instances of RRNs, and on clarifying the connection between the CSPs discussed in \cite{Article_popDyn} and the Network Expansion scheme \cite{Ebenhoh}. Concerning the latter point, we have shown that NE is recovered as a limiting case of the present CSPs, and that our method permits a thorough exploration of its solution space,  much beyond the straightforward computational approaches employed previously. Moreover, after computing the exact phase diagram of NE, we have quantitatively connected the transitions one observes to percolation phenomena. Moving on to the general CSPs, our results for single instances turn out to be in remarkable agreement with the population dynamics study of \cite{Article_popDyn}. As expected for a RRN, the solution space is robust to changes in the availability of nutrients, a feature that is unlikely to be transferred over to more realistic networks (e.g. cellular metabolic networks).

The method presented here can be generalized to include a certain fraction of reversible reactions in a straightforward way, and applicability to more realistic (real) network topologies could mainly be limited by convergence issues. Future work will be exploring this aspect and, more importantly, the emerging picture of the solution space on bacterial metabolic networks.

\acknowledgments

This work is supported by the Italian Research Minister through the FIRB Project No. RBFR086NN1 and by the DREAM Seed Project of the Italian Institute of Technology (IIT). The IIT Platform Computation is gratefully acknowledged.

\bibliographystyle{unsrt}
\bibliography{biblio}

\appendix

\section*{Appendix}

\subsection{Cavity equations}
\label{cavity_equation}
In general a CSP, as Soft-MB or Hard-MB, can be solved efficiently on random networks by the belief propagation algorithm \cite{Pearl_BP} or equivalently by the replica symmetric cavity method \cite{Parisi_cavity}. In this method, the marginal of a variable is computed by creating a ``cavity'' inside the system, removing a subpart of the network. Thus it is possible to obtain a ``cavity marginal'' and then reintroduce the variables removed. Finally the complete marginal of the variables follows directly from the cavity marginals. 

\begin{figure}
  \centering
  \includegraphics[width=8.5cm]{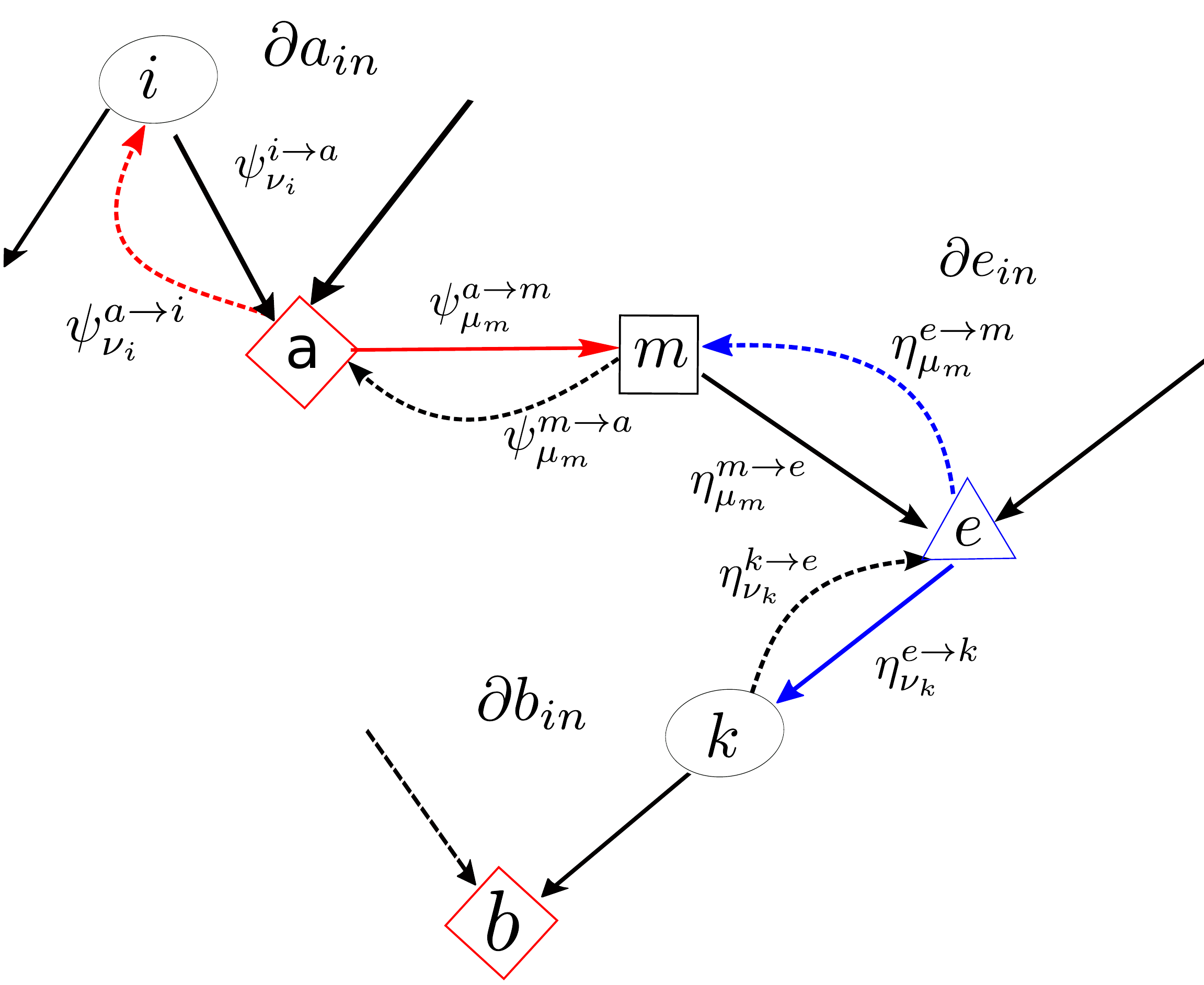}
  \includegraphics[width=8.5cm]{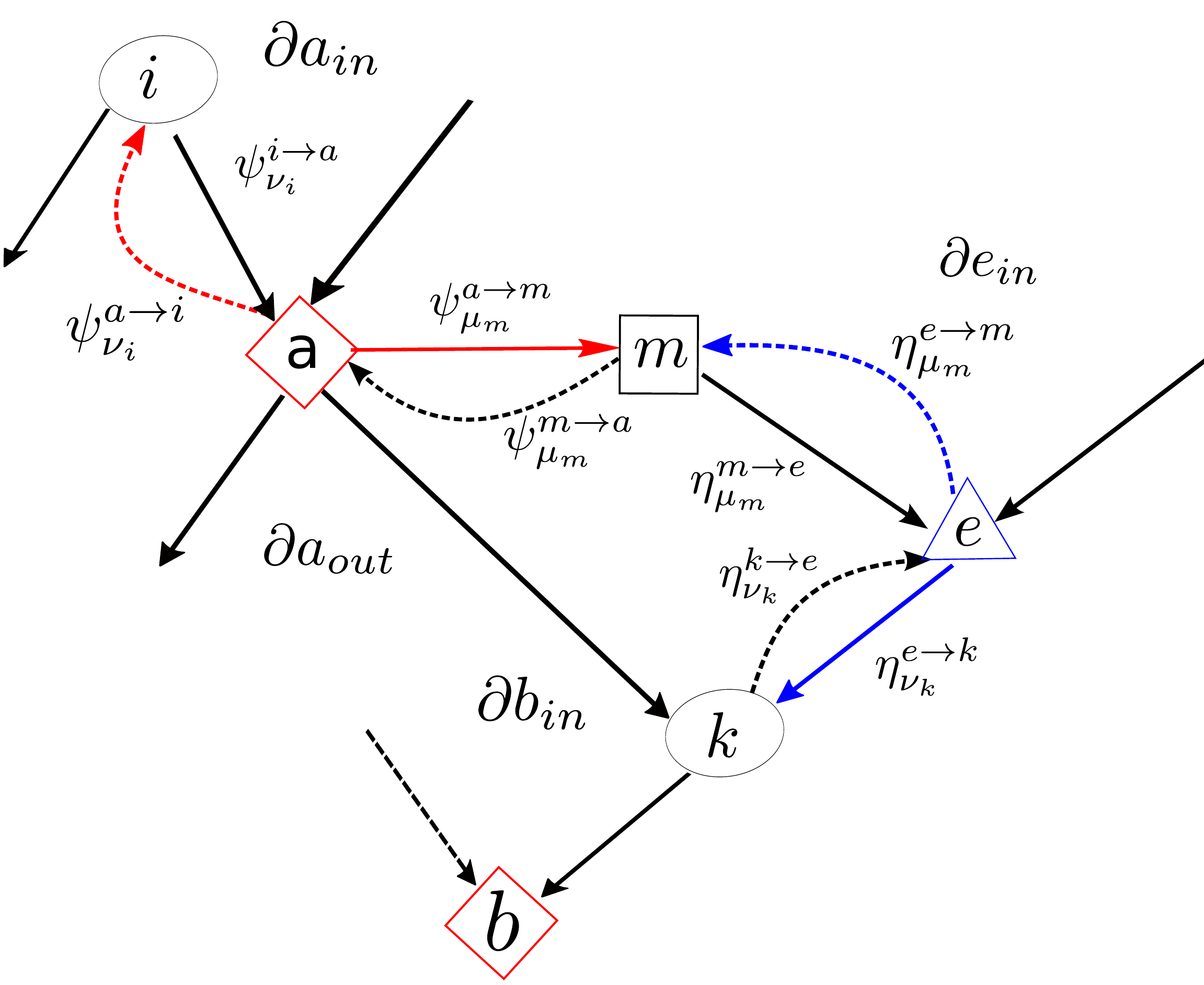}
  \caption{Schema of the cavity method for Soft-MB (left) and Hard-MB (right) constraints.   \label{schema_cavity}}
\end{figure}

In this kind of approach the system is divided between ``variable'' and ``function'' nodes. In the RRN this amounts to add two types of function nodes (constraint $\Gamma$ and $\Delta$) as presented in Figure \ref{schema_cavity}. In the following we will use letters $a,b,..$ for the metabolite constraint and $e,f,..$ for the reaction constraint. Furthermore we introduce the condensed notations: $\partial a^R=\partial a \backslash m$, as the reaction neighbours of the metabolite constraint $a$;  $\partial e^M=\partial e \backslash i$, as the metabolites neighbours of the reaction constraint $e$; $\partial a^R_i$ represents the reaction neighbours of $a$ that are in the same group as $i$ without $i$ and $\partial a^R_{\neg i}$ the reaction neighbours of $a$ that are in the opposite group of $i$.

The resulting equations for the system are (for a full derivation refer to \cite{Article_popDyn}):
\begin{align*}
&  \begin{cases}
    \mess{\psi}{m}{a}_{\mu_m}=\prod\limits_{f \in \neigRea[m]}\mess{\eta}{f}{m}_{\mu_m}/\Z{m}{a} \\ \\
    \mess{\psi}{a}{m}_{\mu_m}=\left[\delta_{\mu_m,0}\prod\limits_{\neigIn}\mess{\psi}{j}{a}_0\left(\prod\limits_{\neigOut}\mess{\psi}{j}{a}_0\right)^{\alpha}+\delta_{\mu_m,1}\left(1-\prod\limits_{\neigIn}\mess{\psi}{j}{a}_0\right)\left(1-\prod\limits_{\neigOut}\mess{\psi}{j}{a}_0\right)^{\alpha}\right]/\Z{a}{m}
  \end{cases}
  \\ \nonumber \\
  &  \Z{a}{m}=\left(1-\prod_{\neigIn}\mess{\psi}{j}{a}_0\right)\left(1-\prod_{\neigOut}\mess{\psi}{j}{a}_0\right)^{\alpha}+\prod\limits_{\neigIn}\mess{\psi}{j}{a}_0\left(\prod\limits_{\neigOut}\mess{\psi}{j}{a}_0\right)^{\alpha}
  \\ \nonumber \\
&  \begin{cases}
    \mess{\psi}{i}{a}_{\nu_i}=\mess{\eta}{e}{i}_{\nu_i}\left(\prod\limits_{b \in \neigMeta[i]_{in} \backslash a}\mess{\psi}{b}{i}_{\nu_i}\right)^{\alpha}\prod\limits_{b \in \neigMeta[i]_{out} \backslash a}\mess{\psi}{b}{i}_{\nu_i}/\Z{i}{a} \\ \\
    \mess{\psi}{a}{i}_{\nu_i}.\Z{a}{i} =\mess{\psi}{m}{a}_0(1-\nu_i)\prod\limits_{\neigIn \backslash i}\mess{\psi}{j}{a}_0\left(\prod\limits_{\neigOut \backslash i}\mess{\psi}{j}{a}_0\right)^{\alpha}+\\ \\+\mess{\psi}{m}{a}_1\left(1-\prod\limits_{\neigOppositeI}\mess{\psi}{j}{a}_0\right)^{\alpha}\left((1-\prod\limits_{\neigGroupI}\mess{\psi}{j}{a}_0)+\nu_i\prod\limits_{\neigGroupI}\mess{\psi}{j}{a}_0\right)
  \end{cases}
  \\ \nonumber \\
&\Z{a}{i}=\mess{\psi}{m}{a}_0\prod\limits_{\neigGroupI}\mess{\psi}{j}{a}_0\left(\prod\limits_{\neigOppositeI}\mess{\psi}{j}{a}_0\right)^{\alpha}+ \mess{\psi}{m}{a}_1 \left(1-\prod\limits_{\neigOppositeI}\mess{\psi}{j}{a}_0\right)^{\alpha}\left(2-\prod\limits_{\neigGroupI}\mess{\psi}{j}{a}_0\right)
\end{align*}
and, for the reaction constraints,
\begin{align*}
&  \begin{cases}
    \mess{\eta}{i}{e}_{\nu_i}=\left(\prod\limits_{b \in \neigMeta[i]_{in}}\mess{\psi}{b}{i}_{\nu_i}\right)^{\alpha}\prod\limits_{b \in \neigMeta[i]_{out}}\mess{\psi}{b}{i}_{\nu_i}/\Z{i}{e}  \\ \\
    \mess{\eta}{e}{i}_{\nu_i}=\left[\delta_{\nu_i,0}+e^{\theta}\delta_{\nu_i,1}\prod\limits_{n\in \neigMeta}\mess{\eta}{n}{e}_1\right]/\Z{e}{i} 
  \end{cases} 
  \\ \nonumber \\
& \Z{e}{i}=1+e^{\theta}\prod_{m \in \neigMeta}\mess{\eta}{m}{e}_1
  \\ \nonumber \\
&  \begin{cases}
    \mess{\eta}{m}{e}_{\mu_m}=\mess{\psi}{a}{m}_{\mu_m}\prod\limits_{f \in \neigRea[m] \backslash e}\mess{\eta}{f}{m}_{\mu_m}/\Z{m}{e} \\ \\
    \mess{\eta}{e}{m}_{\mu_m}=\left[\mess{\eta}{i}{e}_0+e^{\theta}\mess{\eta}{i}{e}_1\mu_m\prod\limits_{n\in \neigMeta \backslash m}\mess{\eta}{n}{e}_1\right]/\Z{e}{m}
  \end{cases}
  \\ \nonumber \\
& \Z{e}{m}=2\mess{\eta}{i}{e}_0+e^{\theta}\mess{\eta}{i}{e}_1\prod\limits_{n\in \neigMeta \backslash m}\mess{\eta}{n}{e}_1
\end{align*}

\begin{figure}[h]
  \centering
  \includegraphics[scale=0.35]{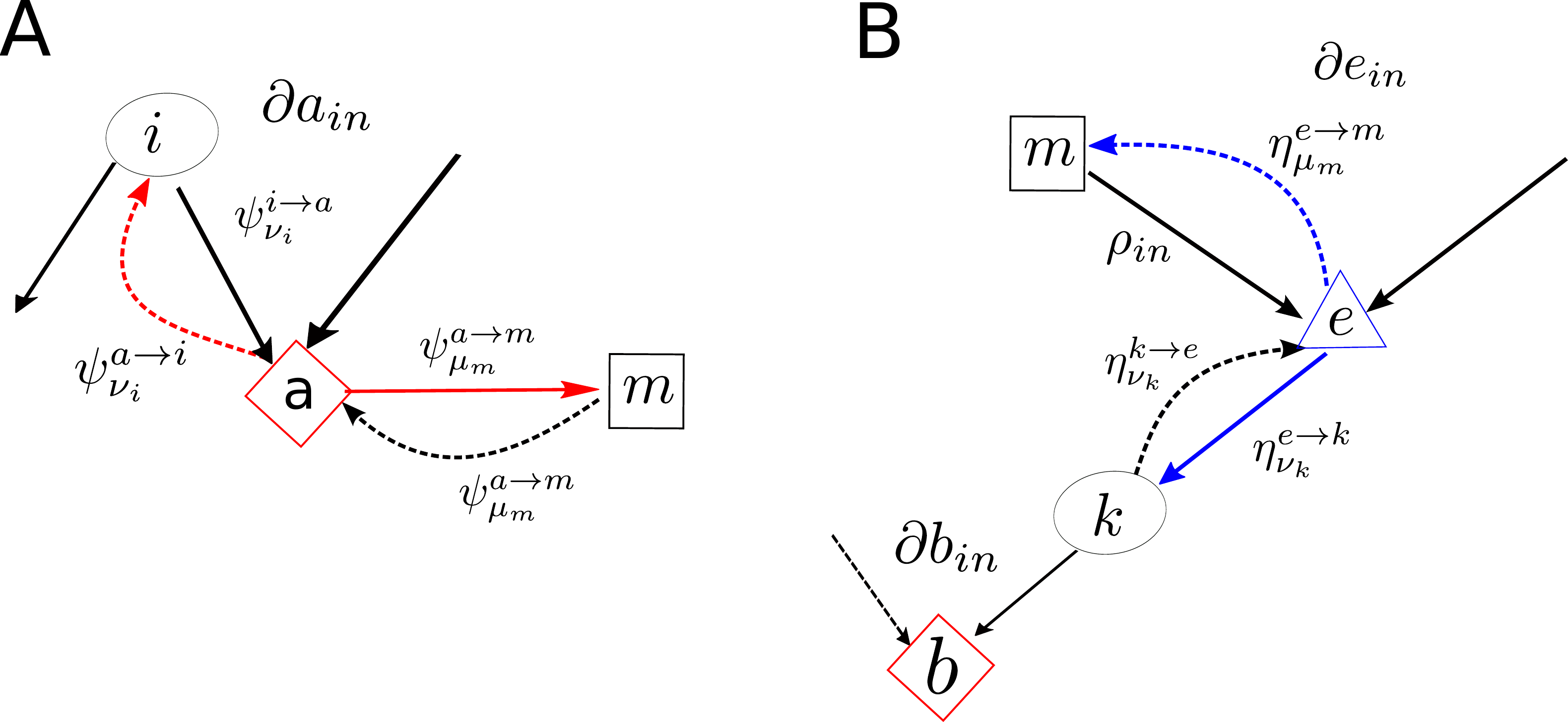}
  \caption{Representation of the external metabolites in our network. \textbf{A} is the product while \textbf{B} is the nutrient.}
  \label{nutrOut_schema}
\end{figure}

It is important to note that nutrients are treated differently from the rest of the network. In fact in the last article we decided that nutrients shouldn't have associated metabolite-constraint while sinks have it (Figure \ref{nutrOut_schema}). Hence looking at constraint (\ref{constraint_rea}), it is immediately clear that in this setting if a nutrient is present, its neighbouring reactions can be either active or not, whereas if nutrient is absent no neighbouring reaction can function. Indeed this is how nutrients are used in real networks.

\subsection{Methods}
\subsubsection{Belief Propagation Algorithm}
\label{BP_explanation}
Belief Propagation is an algorithm for efficient unbiased sampling of the solutions of a set of equations \cite{Pearl_BP}. In a nutshell, in BP it is considered that each variable sends a message to its neighbours. This message represents the belief that the variables has about the state of its neighbours. The outcome of this algorithm is the BP-marginal for variables, $\mu$ and $\nu$.

It is worth noting that while in the complete case many different messages exist between the variables (see Appendix \ref{cavity_equation}), in a Mean Field approximation, the messages are the same for all neighbours and correspond to $\avg{\mu_m}$ and $\avg{\nu_i}$. Nevertheless the functioning of the algorithm is similar in the two cases: first we generate a RRN with a given $q$ and $\lambda$, then we initialize the messages (to a random value or to the last value computed) and we iterate the equations until convergence. Finally for the complete problem (in Mean Field the BP-marginal is equal to the marginal) at convergence it is possible to recover the marginals as:
\begin{align}
& p(\mu_m)=\mess{\psi}{a}{m}_{\mu_m}\prod\limits_{f \in \neigRea[m]}\mess{\eta}{f}{m}_{\mu_m}/Z^m,
\nonumber \\ \label{proba_rel} \\ \nonumber
& p(\nu_i)=\mess{\eta}{e}{i}_{\nu_i}\left(\prod\limits_{b \in \neigMeta[i]_{in}}\mess{\psi}{b}{i}_{\nu_i}\right)^\alpha\prod\limits_{b \in \neigMeta[i]_{out}}\mess{\psi}{b}{i}_{\nu_i}/Z^i,
\end{align}
where:
\begin{align}
& Z^m=\sum_{\mu_m}\mess{\psi}{a}{m}_{\mu_m}\prod\limits_{f \in \neigRea[m]}\mess{\eta}{f}{m}_{\mu_m}.
\nonumber \\ \label{normalizations} \\ \nonumber
& Z^i=\sum_{\nu_i}\mess{\eta}{e}{i}_{\nu_i}\left(\prod\limits_{b \in \neigMeta[i]_{in}}\mess{\psi}{b}{i}_{\nu_i}\right)^\alpha\prod\limits_{b \in \neigMeta[i]_{out}}\mess{\psi}{b}{i}_{\nu_i}.
\end{align}
All networks in this paper have $M=10^4$ while $N=\lambda M / (1+q)$.

The simplest way to sample the solutions is by fixing one of the two free variables remained: $\theta$ or $\rho_{in}$. By changing $\rho_{in}$ we can see how the configuration of the solutions changes when the nutrients have a probability $\rho_{in}$ of functioning. Whereas by changing $\theta$ we can observe what happens if we constrain the system to switch on (or off) the reactions. Each behaviour is interesting to understand how the system is organized. In each case the mean over the metabolites,$\avg{\mu}$, and the reactions, $\avg{\nu}$ ($\avg{x}$ is the average over the measure $P(\mu,\nu)$, (\ref{meas})) has been computed.

\subsubsection{Decimation Procedure}
\label{decimation_explanation}
The BP algorithm is an efficient way for obtaining the {\it probability} that a variable take a certain state. Nevertheless one is generally confronted with the problem of obtaining actual \textit{configurations} of variables that satisfy a CSP. In order to find it, we resorted to a decimation procedure already used with success in other cases \cite{RicciT_Semerjian,AurelienFede}.

In decimation, first BP is run and then the BP marginal is used as the real marginal of the variable, thus setting the variable to $0$ or $1$ \textit{according to the marginal}. Hence during decimation, variables are set one at a time, starting from the most polarized (with BP-marginal near $0$ or $1$) then running BP to make sure that the constraints are satisfied and that no contradiction occurs. This procedure is then iterated until all variables are decimated or until some constraint is violated.

Using this procedure it is thus possible to obtain a Boolean configuration that is a solution of the CSP problem under study. It is important to note that while BP is an unbiased way of sampling the solution space (at least for problems on random graphs), the decimation process is highly dependent on the procedure used to decimate. Nevertheless, if the procedure converges, the configuration found will be a solution of the CSP. Furthermore assuming BP marginals are unbiased for a RRN it is possible to understand whether we are sampling fairly well the solution space with decimation.

In order to reproduce the behaviour already observed in \cite{Article_popDyn}, the algorithm that we used to obtain the results presented in Figures \ref{VN_l1_q05}--\ref{FBA_l3_q08} is an extension of the standard decimation procedure presented above. In our algorithm, for a given $\theta$, first a BP solution is found and stored, then the system is decimated $N_{dec}$ times, each time starting from the same BP solution stored. Finally BP solution for the next $\theta$ is obtained by initializing the messages with the last stored BP solution. For each system under study we applied this procedure following the two protocols ($+\infty \rightarrow -\infty$ or $-\infty \rightarrow +\infty$) presented in \cite{Article_popDyn}. All the results in this article have been obtained with $N_{dec}=5$ for the complete problem and $N_{dec}=10$ for MF.

\end{document}